\documentclass[nocopyright]{cimento}

\usepackage{graphicx,caption}  
\usepackage{wrapfig}
\usepackage{cite}
\usepackage{multicol}
\usepackage{amsmath}
\usepackage{xcolor}
\usepackage{amssymb}
\usepackage{mathtools}
\usepackage{bm}
\usepackage{units}
\usepackage[T1]{fontenc} 
\usepackage{url}
\title{Isospin effects and symmetry energy in nuclear reactions}
\author{Jerzy~\L{}ukasik\thanks{jerzy.lukasik@ifj.edu.pl}
}
\instlist{\inst{} IFJ PAN, PL-31342 Krak\'{o}w, Poland}

\begin{document}

\maketitle

%
%
%

\begin{abstract} Isospin and the symmetry energy are linked within the symmetry
term of the liquid-drop mass formula and within the corresponding term of the
equation of state of asymmetric nuclear matter. The Pauli exclusion principle,
resulting from the fermionic nature of nucleons, leads, in the parabolic
approximation, to a quadratic dependence of the kinetic part of the symmetry
term of the nuclear binding-energy formula on the neutron--proton asymmetry
(N-Z)/A. Charge symmetry ensures the same leading quadratic dependence for the
potential part. Charge invariance further generalizes this dependence, to lowest
order, from T$_3^2$ to T(T+1). The strength of the symmetry term, in contrast,
is determined by the symmetry energy. At sub-saturation densities the symmetry
energy can be constrained by nuclear-structure observables, such as neutron skin
thicknesses, isovector collective modes and the related electric dipole
polarizability, and excitation energies of isobaric analog states. It can also 
be constrained by reaction observables, such as charge exchange and elastic
scattering cross sections, isoscaling and isospin diffusion. At supra-saturation
densities, by contrast, the only means of studying asymmetric nuclear matter
under laboratory conditions are relativistic heavy-ion collisions. These studies
use differential observables such as neutron/proton flows and light isobar or
$\pi^{-}/\pi^{+}$ ratios. Complementary information can be obtained from neutron
star observations including gravitational waves. The latest measurements and
analyses of the stiffness of the symmetry energy will be presented, together
with the recent NA61/SHINE observation of an unexpectedly large isospin-symmetry
violation in kaon production. Attention will also be paid to the challenges of
measuring charged and neutral products of relativistic heavy-ion collisions in
the context of the recent ASY-EOS II experiment at GSI and to the new detectors
used in the setup. \end{abstract}

\section{Introduction, basic concepts, symmetries and their breaking}

In 1932, based on observing nearly equal masses and the same spins of the
proton, p, and the neutron, n,  Heisenberg \cite{heis32} proposed to treat these
particles as two states of the ``nucleon'' and to adopt spin formalism to
describe them. A nucleon has been assigned an {\em isospin} quantum number,
T=$\nicefrac{1}{2}$, with two possible projections T$_{3}$=$\pm \nicefrac{1}{2}$
on the third axis in the isospin space, see e.g. \cite{BohrMottelson}.  
The observation of certain symmetries in the subatomic world led to the
formulation of specific hypotheses regarding nuclear interactions. 

For example, explanation of the mass differences between ``mirror nuclei''
($^{3}$H-$^{3}$He, $^{7}$Li-$^{7}$Be, etc.) largely by electrostatic effects and
the n-p mass difference led to the hypothesis of {\bf charge symmetry} of nuclear
forces. It states that, apart from the Coulomb interaction, the interactions
between like nucleons are the same: {\bf (n-n)=(p-p)}, i.e. they do not change
under the n$\leftrightarrow$p interchange.

Observation of an approximate equality of nuclear interactions in p-p and n-p
scattering (in the $^{1}$S channel), similarities of some level schemes in
neighboring isobars (isobaric analog states, IAS, with the same spin, parity and
isospin), selection and transition rules driven by isospin conservation, etc., led
to the hypothesis of {\bf charge invariance} of nuclear forces. It states that the
strong interactions between n-n, p-p and n-p pairs in the same T=1 state, (i.e.
with the same spin and angular momentum), are the same: {\bf (n-n)=(p-p)=(n-p)},
thus they are invariant under arbitrary rotations in isospin space.

Observation of isospin multiplets of mesons and baryons, isospin driven predictions
for relative strengths of certain reaction channels, led to the extension of
isospin to all hadrons and to the hypothesis of {\bf isospin symmetry} of strong
interactions. It assumes that the up, u, and down, d, quarks interact identically
provided their mass and charge difference can be ignored i.e. {\bf u=d}.

Finally, {\bf flavor symmetry} extends isospin to the strange quark and states that
strong interactions are independent of the quark type, i.e. {\bf u=d=s}.

These symmetries are approximate. Isospin symmetry is broken by the u-d quark mass
difference and by electromagnetic interactions, and the flavor symmetry breaking
is dominated by the larger mass of the s quark. Recently, however, the NA61/SHINE
collaboration reported evidence of isospin-symmetry violation beyond these known
effects with the $4.7\sigma$ significance\cite{shine}. The observation was drawn
from the measured charged-to-neutral kaon ratio $R_{K} \equiv (\langle K^{+}
\rangle + \langle K^{-} \rangle)/(2 \langle K^{0}_{S} \rangle)$ in nearly symmetric
collisions of $^{40}$Ar + $^{45}$Sc at $\sqrt{s_{NN}}$ = 11.9 GeV. $R_{K}$ was
found to exceed unity by (18.4 $\pm$ 6.1)\%. Similar observations were made in many
other experiments, but with smaller precision. To date, there is no generally
accepted explanation of this unexpectedly large effect. 

Since no comparably large violation has been found in observables involving only
the u and d quarks (e.g. in $\pi^{-}/\pi^{+}$ ratios), the effect may be specific
to strangeness. If so, it should not affect observables and symmetries discussed in
this contribution. Rather than providing another comprehensive overview of symmetry
energy studies and constraints (see e.g 
\cite{bali08,bali13,lat14,latt16,baldo16,oert17,sorensen24,cozma25,vesic26} and
references therein), this paper aims to identify direct links between isospin and
symmetry energy, outline key (subjective) milestones and highlight most recent
results and future advances. 

\section{Isospin in nuclear structure and dynamics and its relation to Esym}

In 1935 Weizs\"{a}cker \cite{wei35} proposed a semi-empirical ``liquid drop''
formula for binding energy per nucleon in a nucleus of mass, atomic and neutron
numbers A, Z and N, respectively. It was later supplemented  with the pairing and
shell corrections (see e.g. \cite{BohrMottelson}):
\begin{equation}
-\frac{E_{B}}{A} =  
\underbrace{-a_{V} \mathop{\vphantom{\int}}}_{\text{Volume}} 
+\underbrace{a_{S} \frac{1}{A^{1/3}}\mathop{\vphantom{\int}}}_{\text{Surface}} 
+\underbrace{a_{C} \frac{Z^{2}}{A^{4/3}}\mathop{\vphantom{\int}}}_{\text{Coulomb}}
+\underbrace{a_{A}
\textcolor{black}{\left(\frac{N-Z}{A}\right)^{2}\mathop{\vphantom{\int}}}}_{\text{Symmetry}} 
\pm \underbrace{a_{P} \frac{1}{A^{3/2}}\mathop{\vphantom{\int}}}_{\text{Pairing}}
+  \underbrace{\Delta E_{S}\mathop{\vphantom{\int}}}_{\text{Shell Corr.}}
\label{eq_bv}
\end{equation}
The quadratic neutron-proton-asymmetry dependence of the symmetry term, ST, in
(\ref{eq_bv}) can be derived from the Fermi Gas, FG, model and complies with the
charge symmetry of nuclear forces. It results from the fermionic nature of the
nucleons (Pauli exclusion) and specifically from the Fermi energy dependence on
neutron and proton content. Up to the quadratic term in the n-p asymmetry expansion:
\begin{equation}
\frac{E_{tot}^{FG}}{A} \approx
\underbrace{\overbrace{\frac{3}{5} \frac{\hbar^2}{2 m} \left( \frac{3}{2} \pi^{2} 
\rho_{o}\right )^{2/3}}^{\sim \text{22 MeV}}
{\color{black}{ \left(\frac{\rho}{\rho_{o}} \right)^{2/3}}}}_{\text{\color{black}{FG Symmetric Matter Term}}} + 
\underbrace{
\underbrace{\overbrace{\frac{5}{9} \frac{3}{5} \frac{\hbar^2}{2 m} \left(\frac{3}{2} 
\pi^{2} \rho_{o}\right )^{2/3}}^{\sim \text{12 MeV}} \color{black}{ 
\left(\frac{\rho}{\rho_{o}} \right)^{2/3}}}_
{\text{FG Symmetry Energy (kinetic part)}}
\overbrace{\color{black}{\left( \frac{ N-Z }{A}  \right)^2}}^
{ \text{\color{black}{n-p asymmetry}}}}
_{\color{black}{\text{FG Symmetry Term}}}
\label{eq_fer1}
\end{equation}
\noindent where, $E_{tot}^{FG}/A$ is the total ``Fermi motion'' energy per nucleon
in uniform nuclear matter, i.e. the ``Equation of State'', EoS, of an asymmetric FG
at zero temperature.

The non-interacting FG EoS (\ref{eq_fer1}) is, however, too simple. In
particular, it does not predict neutron stars, NS, heavier than 0.7 M$_\odot$
(see e.g. \cite{oppen39,piek18}) while a $2.35\pm0.11$~M$_\odot$ pulsar (PSR
J0952-0607) has been observed recently \cite{roma26}. Nevertheless, the form of
the FG EoS can be used as a starting point in the search for the true EoS of
asymmetric nuclear matter and in particular for the Symmetry Energy, Esym, a
density dependent function multiplying the square of the n-p asymmetry in the ST
of (\ref{eq_fer1}). In many applications (e.g. transport codes) the Esym is
parametrized by adding to the FG kinetic term a potential part with a power law
dependence on density with the exponent $\gamma$ (\ref{eq_fer2}):
\begin{equation}
E_{sym}(\rho)  =
\underbrace{(\text{12 MeV}) 
{\color{black}{ \left(\frac{\rho}{\rho_{o}}
\right)^{2/3}}}}_{\text{\color{black}{kinetic part}}} + 
\underbrace{((a_{A}^{V}-12) \text{MeV}) \color{black}{\left(\frac{\rho}{\rho_{o}} 
\right)^{\gamma}}}_{\text{potential part}}
\label{eq_fer2}
\end{equation}
It may be viewed as a simplified version of the Skyrme interactions
\cite{dutra12}, with $a_{A}^{V} \equiv E_{sym}(\rho_{o})$ being the volume
contribution to the symmetry parameter $a_{A}$ in eq. (\ref{eq_bv}).

Precise density dependence of Esym is currently a subject of intensive experimental
(e.g. \cite{bali13,oert17,DAN17,RIL19,ABB18,ADH21,RUS16,EST21}) and theoretical
(e.g. \cite{FUC06, GAN16, DRI21}) studies. However, some theoretical and empirical
arguments suggest that the exponent $\gamma$ should follow the kinetic part and be
close to $\nicefrac{2}{3}$ \cite{bali26}. Indeed, the result of \cite{RUS16}
yielded $\gamma$=0.72$\pm$0.19, and more examples can be found in \cite{bali26}.

A general form of the EoS of asymmetric nuclear matter can be obtained by taking a
thermodynamic limit of eq. (\ref{eq_bv}), i.e. requiring that the mass and volume
tend to infinity while keeping the density finite, switching off the Coulomb
interaction and allowing the matter to be compressible. In the parabolic
approximation, this EoS takes the form:
\begin{equation}
E(\rho , \delta) =  E(\rho, 0) + \color{black}{E_{sym}(\rho)} \, 
\color{black}{\delta^{2}}
\label{eq_eos1}
\end{equation}
\noindent with the n-p (or isospin) asymmetry $\delta=\frac{\rho_{n} -
\rho_{p}}{\rho}$ and $\rho_{n},\; \rho_{p},\; \rho$ being the neutron, proton and
total densities, respectively. The $E(\rho , 0)$ and $E_{sym}(\rho)\delta^{2}$ are
the symmetric matter term and ST, and $ E_{sym}(\rho) =E(\rho , 1)-E(\rho , 0)$.
The latter relation implies that Esym can be viewed as an energy cost of converting
all protons in symmetric matter into neutrons.

To facilitate experimental extraction of the density dependence of the EoS, both
the symmetric matter term and Esym are usually expanded in Taylor
series around the saturation density $\rho_{o}$. Setting $x=\frac{\rho -
\rho_{o}}{3\rho_{o}}$, the expansions up to the cubic term yield:
\begin{equation}
E(\rho, 0) \approx -a_{V} 
+ \frac{K}{2} x^2 
+ \frac{Q}{6} x^3 
, \;\;\;\;\;\;
 E_{sym}(\rho) \approx  E_{sym}(\rho_o) 
+ L x
+ \frac{K_{sym}}{2} x^2
+ \frac{Q_{sym}}{6} x^3
\label{eq_eosx}
\end{equation}
While the parameters $a_{V}$ (volume term in eq. (\ref{eq_bv})), $K$
(incompressibility of symmetric matter) and $\rho_{o}$ are relatively well known
($a_{V}=16$ MeV, $K=230\pm30$ MeV, $\rho_{o}=0.17\pm0.03 \text{ fm}^{-3}$
\cite{dutra12}), the parameters characterizing the density dependence of Esym
($E_{sym}(\rho_o),\;L,\;K_{sym},\;Q_{sym}$) and correlations among them are
still a subject of intensive experimental research. It is worthwhile to mention
that using the Esym expansion of \eqref{eq_eosx}, assuming \eqref{eq_fer2} and
the Siemens' $\rho^{2/3}$ scaling \cite{bali26}, one can expect the following
correlations: $L=2E_{sym}(\rho_o)$, $K_{sym}=-2E_{sym}(\rho_o)$,
$Q_{sym}=8E_{sym}(\rho_o)$.

The interplay of isospin, T, and Esym becomes apparent on a closer look at the
ST of eq. (\ref{eq_bv}). Charge invariance, i.e. invariance under rotations in
isospin space is a more general property of nuclear interactions than charge
symmetry. It requires the symmetry term to depend on $\text{\bf{T}}^{2}$, with
eigenvalue T(T+1), rather than on its T$_3$ component alone:
\begin{equation}
\text{ST} \; \equiv \; a_{A} \left(\frac{N-Z}{A}\right)^{2}\mathop{\vphantom{\int}}
 = \underbrace{4 a_{A} \left(\frac{\text{T}_{3}}{A}\right)^{2}\mathop{\vphantom{\int}}}
 _{\text{T}_{3}=(N-Z)/2} 
 = \underbrace{4 a_{A} \left(\frac{\text{T}}{A}\right)^{2}\mathop{\vphantom{\int}}}
 _{\substack{\text{for most ground} \\ \text{states T}=|\text{T}_{3}|}} 
 \;\longrightarrow\; \underbrace{4 a_{A} \frac{\text{T(T+1)}}{A^{2}}\mathop{\vphantom{\int}}}
 _{\substack{\text{charge-invariant} \\ \text{form}}} 
\label{eq_wig}
\end{equation}
\noindent The last step expresses the ST through the total isospin, and its part
linear in T contributes to the Wigner energy \cite{wign37}. This form is supported
by excitation energies of the IAS\cite{jaen03,kirs08,dani14}. It also justifies the
interchangeable use of the terms neutron-proton asymmetry and isospin asymmetry for
most of the ground states.

\begin{figure} [!h]	
 \centering
 \begin{minipage}[c]{0.4\linewidth}
  \centering
  \includegraphics*[width=.99\linewidth]{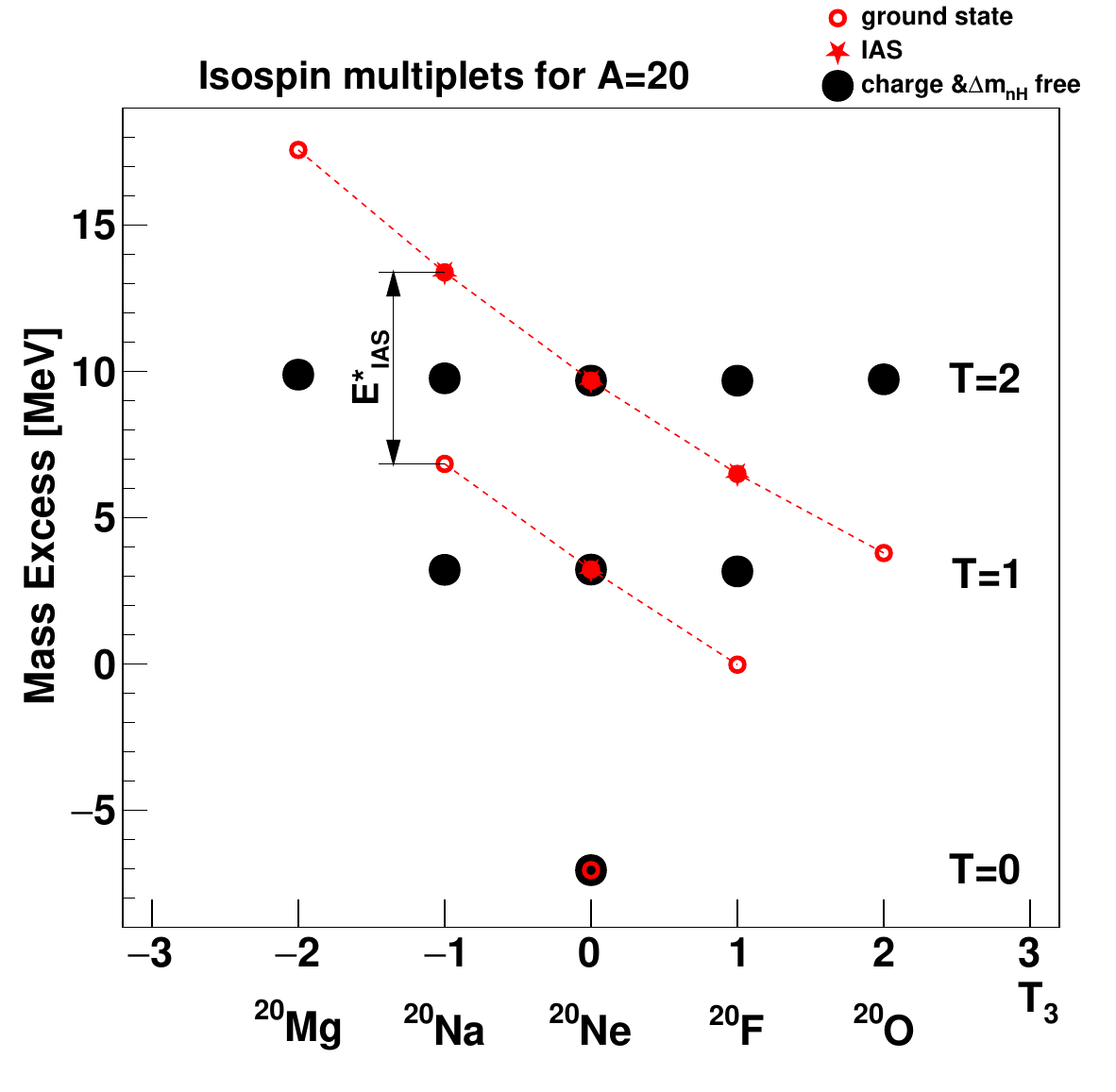}
 \end{minipage}%
 \hfill
 \begin{minipage}[c]{0.58\linewidth}
 
  \caption{Mass excess vs T$_{3}$ for isospin singlet (T=0), triplet (T=1) and
quintet (T=2) of A=20 isobars. Small open circles and stars show experimental
ground state and IAS values, respectively. Filled circles show ``bare strong'' mass
excesses after subtracting the Coulomb energy and the contribution of the
n--hydrogen mass difference, relative to the N=Z member. The vertical arrow marks
the excitation energy of the T=2 IAS in $^{20}$Na. Up to pairing and shell
corrections, this energy is approximately proportional to the symmetry coefficient
$a_{A}$ in eqs. (\ref{eq_bv}) and (\ref{eq_ias}). Note the much larger gap between
the T=0 and T=1 states (see text).}

\label{fig_multi} \end{minipage} \end{figure}

The isospin formalism allows one to assign T to measured ground and excited
states of nuclei and to identify analog levels within isobaric chains. As an
example, Fig. \ref{fig_multi} shows the mass excesses, taken from the
compilation \cite{bri98}, for the ground states (small open circles) and IAS
(stars) of A=20 isobars with T=0, 1 and 2 as a function of T$_{3}$. 

The measured values for T=1 and 2 (symbols connected by lines) are split and
depend almost linearly on T$_3$, as described by the isobaric multiplet mass
equation (IMME\cite{bri98}). This is because the Coulomb interaction and the
neutron--hydrogen mass difference break isospin symmetry. Once these
contributions are subtracted (filled circles, with corrections calculated
relative to the T$_3$=0 isobar following~\cite{wig41}), the members of each
multiplet become nearly degenerate, as expected from charge invariance of
nuclear forces. The energy gap between the T=2 and T=1 multiplets, about
$6.5$~MeV, is then the same for all T$_3$, and reflects the isospin dependence
of the nuclear energy, i.e. the symmetry energy. The gap between the T=1 and T=0
states, about $10.3$~MeV, is much larger. The difference reflects the additional
binding of the even--even N=Z ground state due to the Wigner energy and pairing.
These isobaric multiplets are the nuclear analogs to the isospin multiplets in
the baryon and meson spectra.

The simplest relation between the symmetry coefficient $a_A$ of
eq.~(\ref{eq_bv}) and isospin follows from the excitation energy of an IAS with
isospin T+$\Delta$T above the ground state of the same nucleus (see arrow in
Fig. \ref{fig_multi}):
\begin{equation}
a_{A}\; \approx \;A \; \frac{E_{\text{IAS}}^{*}}
{4 \;\Delta T(2T+\Delta T+1)}\; = \;A \; \frac{E(A,T+\Delta T)-E(A,T)}
{4 \;\Delta T(2T+\Delta T+1)}
\label{eq_ias}
\end{equation}
\noindent where $E(A,T)$ is the energy of the lowest state with isospin T in the
nucleus (eq. (\ref{eq_ias}) corresponds to eq. (2-18) of \cite{BohrMottelson} for
$\Delta T=1$, $T^2$ form of ST and $b_{sym}=2 a_{A}$). Since both states have the
same Z, the Coulomb energy and the neutron--hydrogen mass difference largely
cancel. The approximation neglects pairing and shell effects. The resulting $a_A$
is an effective coefficient for a given nucleus, including surface effects.
Transitions from T=0 ground states of even--even N=Z nuclei must take into account
Wigner energy, pairing and shell corrections.

Different neutron-proton content of a nucleus, through $\text{T}_{3}$, and its
isospin T, which can differ from $|\text{T}_{3}|$ for excited states such as the
IAS, lead, through Esym, to observable differences between nuclei. In nuclear
structure, Esym affects the excitation energies of IAS \cite{dani14},
neutron-proton distributions and skins \cite{chen10}, masses\cite{kor10}, electric
dipole polarizability~\cite{roca13}, giant~\cite{trip08} and pygmy \cite{carb10}
dipole resonances. In nuclear reactions, the density and momentum dependence of the
symmetry potential governs isospin dynamics. This ``isospin transport'' can be
related to such phenomena and observables as isospin drift and
diffusion\cite{pian21,ciam25}, transparency\cite{lefe25},
fractionation/distillation\cite{xu00}, isoscaling\cite{bot02,tsan01,ogul11},
flows\cite{RUS16,cyts24}, $\pi^{-}/\pi^{+}$ ratios\cite{EST21}. 

While the structure related ,``static'', isospin effects impact the Esym mostly
below  $\rho_{o}$ (roughly between 0.4-0.8 $\rho_{o}$ for  10$\leq$A$\leq$240), the
``dynamic'' isospin effects can be studied in heavy ion collisions up to 2-3
$\rho_{o}$ at relativistic energies.

In nuclear collisions, the isospin-sensitive observables are primarily shaped by
``static'' (initial state) isospin effects related to the structure of the
projectile and target (N/Z, neutron--proton density distributions, skins) and by
dynamically induced gradients of the isovector part of the mean field.
Additional contributions arise from in-medium isospin-dependent nucleon-nucleon
cross sections, Coulomb interactions, clustering, and secondary decays.

For example, in peripheral collisions of heavy nuclei at intermediate energies,
the ``neck'' formed during the collision is neutron rich not only because of the
static neutron skins of the projectile and the target, but also due to the
dynamically induced gradients of the symmetry potential which exert forces of
opposite sign on neutrons and protons. The resulting isospin currents are
governed by the symmetry energy, its density slope, and the local density and
isospin asymmetry gradients (see eq. \eqref{eq_drift} based on
\cite{baran05,rizzo08}). Density gradients lead to isospin migration (drift) of
neutrons towards the dilute neck, at a rate set by $\partial E_{\rm
sym}/\partial\rho$. Asymmetry gradients lead to isospin diffusion, at a rate set
by $E_{\rm sym}$. Both drive the system towards chemical equilibrium, with
uniform difference of chemical potentials: $\mu_n-\mu_p = 4E_{\rm
sym}(\rho)\,\delta$. Because $E_{\rm sym}$ is smaller at low density, this
equilibration leads to a neutron-rich neck.
\begin{equation}
\underbrace{
\mathbf{j}_{n}-\mathbf{j}_{p} \mathop{\vphantom{\int}}}
_{\text{isovector current}}
\propto \;\,
-\;\underbrace{
\rho\,\delta \;\overbrace{\frac{\partial E_{\text{sym}}}{\partial \rho} \mathop{\vphantom{\int}}}^{\sim L}
\overbrace{\nabla \rho \mathop{\vphantom{\int}}}^{\substack{\text{density} \\ \text{gradient}}}
 \mathop{\vphantom{\int}}}_{\text{drift}}
\;-\;
\underbrace{
\rho \;E_{\text{sym}}
\overbrace{\nabla \delta \mathop{\vphantom{\int}}}^{\substack{\text{isospin} \\ \text{gradient}}}
 \mathop{\vphantom{\int}}}_{\text{diffusion}}
\label{eq_drift}
\end{equation}

\section{Some recent results and milestones}

\begin{figure} [!htb]	
 \centering
 \begin{minipage}[c]{0.4\linewidth}
  \centering
  \includegraphics*[width=.99\linewidth]{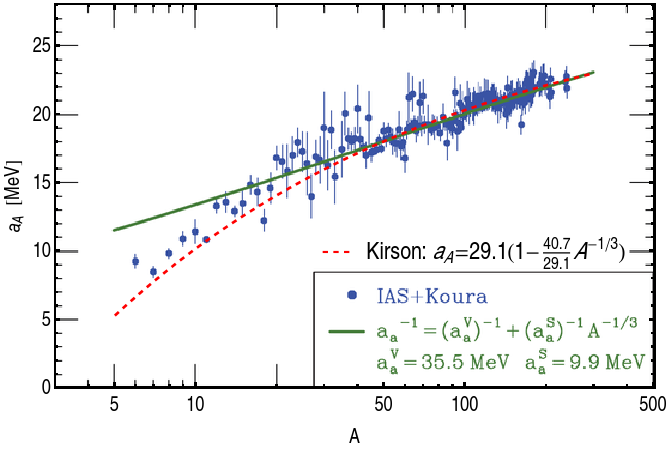}
 \end{minipage}%
 \hfill
 \begin{minipage}[c]{0.58\linewidth}
 
  \caption{Mass dependence of the symmetry energy coefficient $a_{A}(A)$ extracted
from IAS after applying shell corrections. The solid line represents a fit at A
$\geq$ 30 assuming a non-linear combination of volume and surface symmetry terms.
Base Figure taken from \cite{dani14}. The superposed dashed line represents a
result of \cite{kirs08} assuming a linear combination of the volume and surface
symmetry coefficients.}

\label{fig_pd} 
\end{minipage} 
\end{figure}

Systematic studies of symmetry energy coefficients for different nuclei based on
IAS were done e.g. in \cite{jaen03,kirs08,dani14}. The analysis of \cite{dani14} is
more advanced than suggested by eq. (\ref{eq_ias}) as it takes into account shell
corrections. It confirmed that ST is $\propto$~T(T+1) and thus that the
charge invariance is a good approximation. It also provided values of volume and
surface symmetry energy parameters $a_{A}^{V} \text{ and } a_{A}^{S}$ and arrived
at the following constraints: $\mathbf{a_{A}^{V} = 30.2-33.7 \textbf{ MeV and } L =
35-70\textbf{ MeV}}$. The extracted symmetry energy coefficients $a_{A}$ for nuclei
up to A=240 are presented in Fig.~\ref{fig_pd}.
The superposed dashed line represents the result of \cite{kirs08} assuming a linear
combination of the volume and surface symmetry coefficients. It was obtained from
the fits to the double-beta decay Q values. Description of the IAS data is also
remarkable in this case. 

Charge invariance of nuclear forces requires  the nucleon--nucleus interaction to
depend on an isospin scalar. Together with the symmetry term of the nuclear binding
energy, this implies the existence of a symmetry (Lane) potential~\cite{lane62}:
\begin{equation}
U(\mathbf{r}) = U_{0}(\mathbf{r})
 + 4\,\frac{\mathbf{t}\cdot\mathbf{T}}{\text{A}}\,U_{1}(\mathbf{r}),
\label{eq_lane}
\end{equation}
\noindent where $\mathbf{t}\cdot\mathbf{T}$ is the scalar product of the
isospins of the nucleon ($t=\tfrac12$) and of the target nucleus with
mass number A, and $U_{0}$ and $U_{1}$ are the isoscalar and isovector
(symmetry) optical potentials, respectively. 

The influence of isospin on nuclear dynamics through the Lane potential can be
studied in nucleon-nucleus scattering with and without charge exchange. In case of
charge exchange, the target nucleus is excited to its IAS. Analysis of such
scattering data was performed in \cite{DAN17}. After improving the parameters of
the isoscalar and isovector potentials by fitting the model angular distributions
of (p,p), (n,n) and (p,n) reactions to the experimental data, new sharp constraints
were obtained: $\mathbf{a_{A}^{V} =}$ {\bf33.5-36.4 MeV and L=70-101 MeV}.
%
%
%
%

The isospin effects in dynamics can be observed during the collision of
projectile and target with different neutron content at intermediate energies.
The Esym and its density slope together with isospin and density gradients in
the interaction region will lead to isospin drift and diffusion of nucleons
towards chemical equilibrium in the ``spectators'' (see eq. (\ref{eq_drift})).
The rate of the diffusion will depend on the collision energy (interaction time,
attained densities), centrality of the collision, difference in the neutron
content of the colliding isotopes and on the density dependence of the symmetry
energy.

\begin{figure} [!htb]	
 \centering
 \begin{minipage}[c]{0.4\linewidth}
  \centering
  \includegraphics*[width=.99\linewidth]{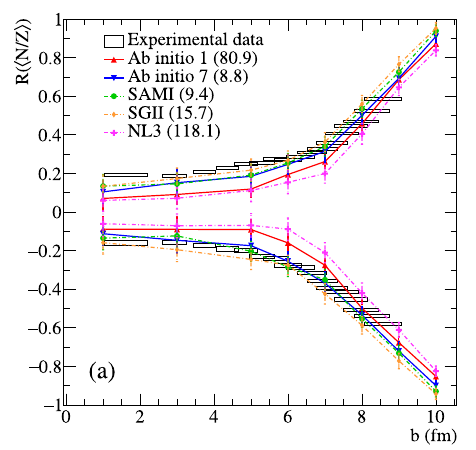}
 \end{minipage}%
 \hfill
 \begin{minipage}[c]{0.55\linewidth}
 
  \caption{ Impact parameter dependence of the isospin transport ratio R obtained
from the $\langle N/Z \rangle$ of quasi-projectile for $^{64}$Ni+$^{58}$Ni and
$^{58}$Ni+$^{64}$Ni reactions at 32 MeV/nucleon. The experimental results  shown as
open rectangles delimiting the uncertainties, are compared with BUU@VECC-McGill
model calculations with equations of state specified in the legend. Figure taken
from Ref. \cite{ciam25}.}

\label{fig_ciampi} 
\end{minipage} 
\end{figure}

The idea of measuring the isospin imbalance ratio \cite{Rami00}, (see eq.
(\ref{eq_rami})), has been used in a recent analysis of the INDRA-FAZIA data on
mean N/Z content of projectile like fragments from  $^{58,64}$Ni+$^{58,64}$Ni
combinations at 32 AMeV for different centralities \cite{ciam25,mall26}.
\begin{equation}
R \left( \left\langle \frac{N}{Z} \right\rangle \right)_{\downarrow}^{\uparrow} =
\frac{ 2 \left\langle \frac{N}{Z} \right\rangle_{58+64}^{64+58} - \left\langle
\frac{N}{Z} \right\rangle_{64+64}^{64+64} - \left\langle \frac{N}{Z}
\right\rangle_{58+58}^{58+58} }{ \left\langle \frac{N}{Z}
\right\rangle_{64+64}^{64+64} - \left\langle \frac{N}{Z}
\right\rangle_{58+58}^{58+58} }
\label{eq_rami}
\end{equation}
The up(down)-arrows in (\ref{eq_rami}) refer to the upper (lower) combinations of
projectiles and targets specified in the numerator and denominator components.
Figure \ref{fig_ciampi} shows the evolution of the measured imbalance ratios with
the centrality of the collision. The best model description of the data has been
obtained for the EoS with \textbf{Esym = 28.53 MeV and L = 48.10 MeV} (blue line in
Fig. \ref{fig_ciampi} for Ab initio 7 parameters).

\begin{figure} [!htb]	
 \centering
 \begin{minipage}[c]{0.4\linewidth}
  \centering
  \includegraphics*[width=.99\linewidth]{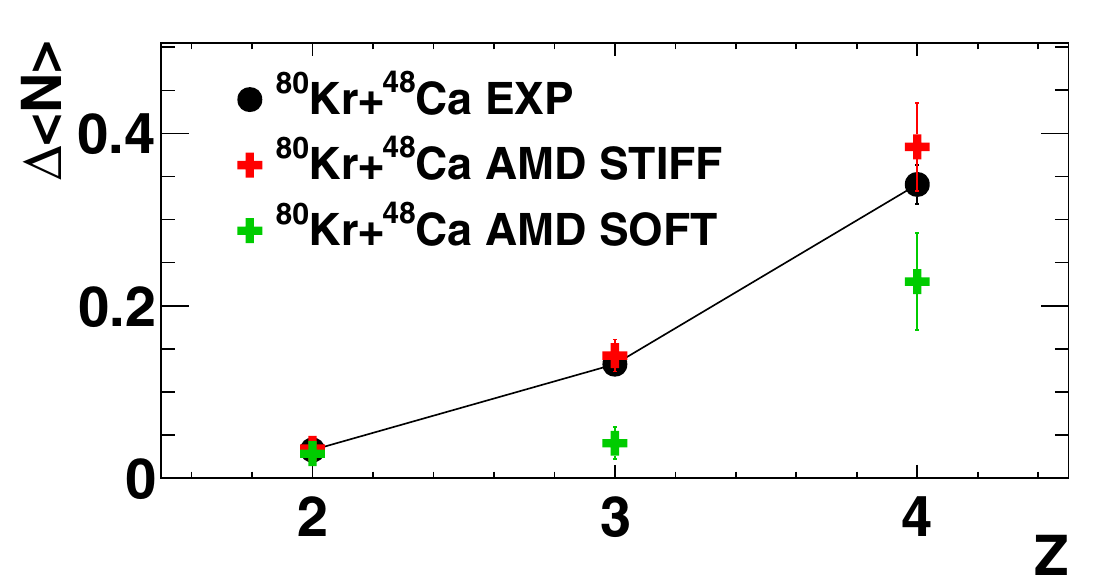}
 \end{minipage}%
 \hfill
 \begin{minipage}[c]{0.55\linewidth}
 
  \caption{Difference between $\langle N \rangle$ of fragments emitted backward and
forward w.r.t. quasi-projectile as a function of their Z for the system  $^{80}$Kr
+ $^{48}$Ca at 35 AMeV. Figure taken from Ref. \cite{pian21}.}

\label{fig_piantelli} 
\end{minipage} 
\end{figure}

Another milestone result in the combined, static isospin effects via the neutron
density distributions, and dynamic isospin migration, on the n/p content of
heavy ion reaction products can be found in \cite{pian21}. The analysis of 
peripheral $^{80}$Kr + $^{40,48}$Ca reactions at 35 MeV/nucleon showed that the
isospin effects are small and difficult to measure, and that special, Esym
sensitive, observables are needed to observe them. Such an observable is
presented in Fig. \ref{fig_piantelli}.
The analysis of fragments emitted in the projectile region showed that the 
secondary emissions wash out the  sensitivity of $\langle N \rangle$/Z or
$\sigma_{N}(Z)$ to the stiffness of Esym. 
Some sensitivity has  been found for the difference  between $\langle N \rangle$
of fragments emitted backward - forward w.r.t. quasi-projectile (Fig.
\ref{fig_piantelli}). Comparison with the AMD model predictions showed a
preference for a rather stiff symmetry energy: \textbf{L $\simeq$ 108~MeV}.
Nevertheless, as shown in \cite{ogul23}, the main experimental results of
\cite{pian21} can also be reproduced within a statistical multifragmentation
model, SMM, after careful adjustment of the size and composition of the
projectile and mid-velocity sources and the charge dependence of the liquid drop
symmetry energy parameter.

Isotopic transparency effects in central $^{124,129}$Xe+$^{112,124}$Sn
collisions at 100 AMeV (INDRA@GSI) have been studied recently in \cite{lefe25}.
The analysis allowed to extract the chemical potentials of n and p (see also
\cite{bonn25}). An incomplete isospin equilibration between the projectile and
target for the reaction products at forward angles has been observed, the
heavier the fragment, the stronger the effect. The analysis also allowed to
deduce a moderate transparency for hydrogen isotopes, whereas for heavier
fragmentation products, with atomic number Z $\geq$ 3, a high transparency
exceeding 50\% was observed. An anomalously large transparency was found for
alpha particles.

Symmetry energy effects have also been recently studied at RIKEN through the near
threshold pion production in $^{108,132}$Sn + $^{112,124}$Sn collisions at 270 AMeV
by the S$\pi$RIT collaboration \cite{EST21}. Interpretation of the measured
spectral pion ratios within the dcQMD model allowed to derive constraints on the
Esym parameters: \textbf{32.5 $<$ Esym$\bm{(\rho_{o})}$ $<$ 38.1 MeV and 42 $<$ L
$<$ 117 MeV} and on the scaled difference between neutron and proton 
effective masses: $\Delta m_{np}^{*} = 0.004 \delta$.

The same data were also analyzed in the context of a coalescence invariant flow of
light charged particles and of their stopping \cite{cyts24}. Experimental
observables have been interpreted in terms of the ImQMD-Sky model within Bayesian
framework yielding the constraints: \textbf{Esym$\bm{(\rho_{o})}$ = 35$\bm{\pm}$3
MeV and L = 83$\bm{\pm}$18  MeV} and $\Delta m_{np}^{*} =
(-0.07\pm0.07)\; \delta$, in a very good agreement with the above pion
analysis. 

The S$\pi$RIT data (neutron-rich $^{132}$Sn + $^{124}$Sn and neutron-deficient
$^{108}$Sn + $^{112}$Sn systems at 270 AMeV) were also analyzed in terms of
directed, v$_1$, and elliptic, v$_2$, flow of hydrogen and helium isotopes, with
the main focus on the momentum dependence of the nuclear mean field
\cite{mizuki25}. The data were interpreted within the AMD model with dynamical
cluster production and momentum dependent mean-field. The analysis of the two
systems indicated a slightly larger v$_2$ strength in $^{132}$Sn + $^{124}$Sn,
suggesting a stronger repulsive potential in the neutron-rich system. Additionally,
a comparison of AMD calculations at L = 46 MeV and L = 108 MeV revealed that the
density dependence of the symmetry energy has a relatively modest impact
(approximately 10\%) on v$_1$, while the particle mass dependence of v$_2$ in
$^{108}$Sn + $^{112}$Sn showed a more pronounced effect (about 20\%). The effect of
the momentum dependence of mean field on flow was found to dominate over the
symmetry energy effect.

Measurements of the total neutron removal cross sections in relativistic 
collisions of neutron rich heavy projectiles with hydrogen or carbon targets has
been proposed as a new method to derive precise ($\pm10$ MeV) constraints on L
\cite{aum17}. First measurements of this type,  $^{12}$C and $^{120}$Sn +$^{12}$C
\cite{ponn25} and $^{124-134}$Sn + C, CH$_2$ and Pb\cite{kuda24}, were performed
within the R3B setup at GSI. Preliminary analyses suggest promising results on
neutron removal, charge changing and total reaction cross sections and their impact
on Esym. 

A similar measurement of the total and charge changing cross sections for
neutron-rich $^{63-80}$Cu isotopes was performed at RIKEN \cite{taka26}. The
impressive data set over 18 isotopes of Cu are expected to provide precise
constraints for models aiming at extracting their neutron skin thicknesses and
the L parameter.


\section{New high energy experiments, new detectors}

The symmetry energy quest continues. New experiments are planned, e.g. S$\pi$RIT
III @ RIKEN in 2027-28 for measuring the $^{208,186}$Pb+$^{208}$Pb collisions at
250 AMeV and ASY-EOS III in the new R3B High-Energy Cave at FAIR in 2029-31 with
completed and possibly split NeuLAND. Radioactive beams (e.g. $^{100-132}$Sn)
between 200 and 1000 AMeV with higher intensity ($\times$2 compared to ASY-EOS
II) are planned to be used for measuring pions with NeuLAND and measuring
projectile and target combinations to extract isospin imbalance ratios. New
detectors (and facilities) are being built and new data are being analyzed.

\begin{figure} [!htb]	
 \centering
 \begin{minipage}[c]{0.49\linewidth}
  \centering
  \includegraphics*[width=.99\linewidth]{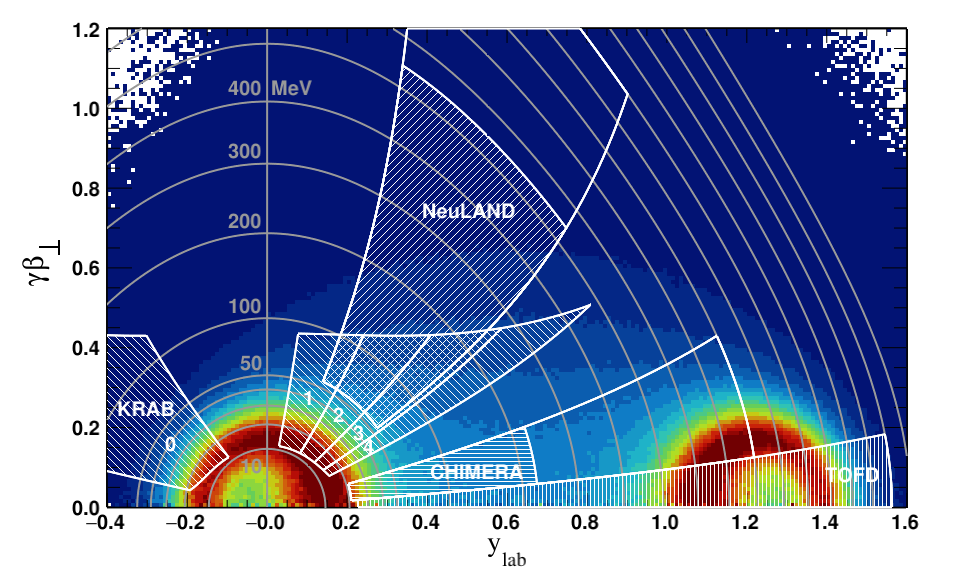}
 \end{minipage}%
 \hfill
 \begin{minipage}[c]{0.51\linewidth}
 
  \caption{Acceptance of the ASY-EOS II detectors for protons. Background:
UrQMD-3.4 prediction of proton invariant cross sections from the 800 AMeV Au+Au
collisions, including clustering and Coulomb afterburner calculations. Realistic
KRAB thresholds for more than 17k photons produced in the scintillating fibers
were applied (see the 5 regions corresponding to KRAB rings 0-4.}

\label{fig_accept} 
\end{minipage} 
\end{figure}

\begin{figure} [!htb]	
 \centering
 \begin{minipage}[c]{0.49\linewidth}
  \centering
  \includegraphics*[width=.93\linewidth]{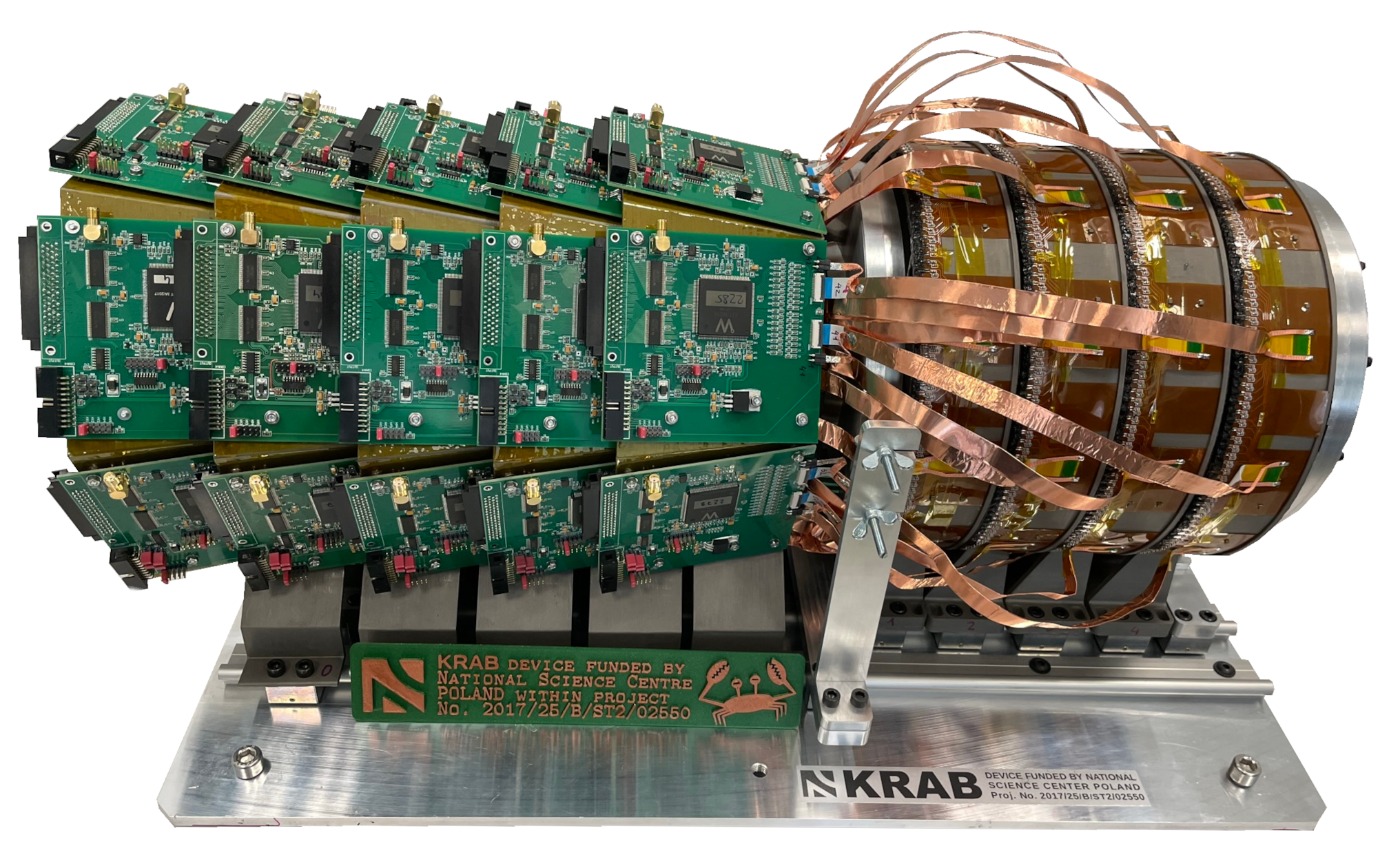}
 \end{minipage}%
 \hfill
 \begin{minipage}[c]{0.51\linewidth}
 
  \caption{KRAB detector with its four forward rings visible. The backward ring is
  hidden beneath the green CITIROC boards. A vertical aluminum bracket supports an
  aluminum flange that houses a remotely controlled four-slot target wheel. The
  copper shielded flat cables connect the SiPMs with CITIROC ASICs. A base with
  rails allows for precise adjustment of the ring position.}

\label{fig_krab} 
\end{minipage} 
\end{figure}

In particular, analysis of a new set of data from 2025 ASY-EOS II campaign
measuring the Au+Au collisions @ 280, 400, 600 and 1000 AMeV is ongoing (see the
contributions of F. Risitano and E. Gambera to these Proceedings). The experiment
is expected to provide improved resolutions and tighter constraints on L,
E$_{sym}(\rho_o)$ and possibly K$_{sym}$. The setup consisted of the NeuLAND, KRAB
and CHIMERA multi-detectors, along with the two TOFD time of flight walls. The
acceptance of the setup for protons is depicted in Fig. \ref{fig_accept}.
The white contours indicate the lower and upper energy thresholds of the detectors
and their angular range. The dashed areas for NeuLAND and CHIMERA mark regions of
possible mass identification for stopped particles. The NeuLAND and CHIMERA
detectors focus on detection of the mid-rapidity emissions, while the TOFD is
focused on detection of projectile fragmentation products. The KRAB detector is
triggering mostly on target emissions. 
The new KRAB detector, depicted in Fig. \ref{fig_krab}, was succesfully used in the
ASY-EOS II experiment to provide fast multiplicity trigger and the data necessary
for centrality and reaction plane estimates.
This 736-channel plastic, passing-through detector enabled also a relatively
undisturbed measurement of neutrons and protons in the NeuLAND neutron detector.
The precise neutron and proton detection was a key requirement of the experiment
aimed at providing symmetry energy constraints from the n/p flow excitation
function. KRAB also allowed to face two major challenges of experiments performed
``in air'': $\delta$-electrons and contamination by the off-target reactions
(mainly on air and on up-stream material). Triggering by $\delta$-electrons was
eliminated by optimally setting the discriminator thresholds. Off-line elimination
of off-target events was achieved by inspection of the forward/backward
multiplicities in KRAB rings, identification of tracks passing through the backward
and forward rings and anti-correlations of reaction planes in KRAB and CHIMERA
detectors.

KRAB uses the 4$\times$4 mm$^2$ plastic scintillation fiber segments (BCF-10)
within an innovative thin mechanical structure 3D-printed from the ABS. This
resulted in a relatively transparent transmission detector. The photons
generated in the scintillators were counted by multi-pixel silicon
photomultipliers (SiPM, SensL MicroFJ-30035) of an area of 3$\times$3 mm$^2$.
Four larger-diameter forward rings and a smaller backward ring (hidden behind
the electronic boards in the photo) cover polar angles from about 30$^o$ to
165$^o$ with a geometric efficiency of approx. 85\% within its angular range. Pulses from the SiPMs were
processed by 23 32-channel CITIROC 1A ASICS, providing logic signals for pulses
meeting the required amplitude criteria. The detection system also included the
TriggerBox and ControlBox electronic devices as well as a control software. The
TriggerBox, based on 30 FPGA matrices, generated a bit pattern containing
information about fired segments and a trigger signal when a pre-defined
multiplicity threshold was exceeded. The decision to generate the trigger was
made very fast, in about 50-100 ns. Together with the serialization and
transmission times, this resulted in about 0.9 $\mu$s intrinsic dead time
introduced by this triggering detector. The bit patterns of the fired segments
were sent to the V2495 logic module, which was the core component of the
associated VME data acquisition system. The ControlBox, built on the basis of
compact SiPM A7585D power supplies, enabled remote setting of individual
voltages on sensors and of gains and discriminator thresholds on the CITIROC
chips, as well as viewing the selected channels. An additional component of the
system was also a "helium sleeve" with a remotely controlled "target wheel" for
changing the targets placed in between the backward and the first forward ring.
The detector performed very well during the experiment.
\begin{figure} [!htb]	
  \centering
  \includegraphics*[width=.9\linewidth]{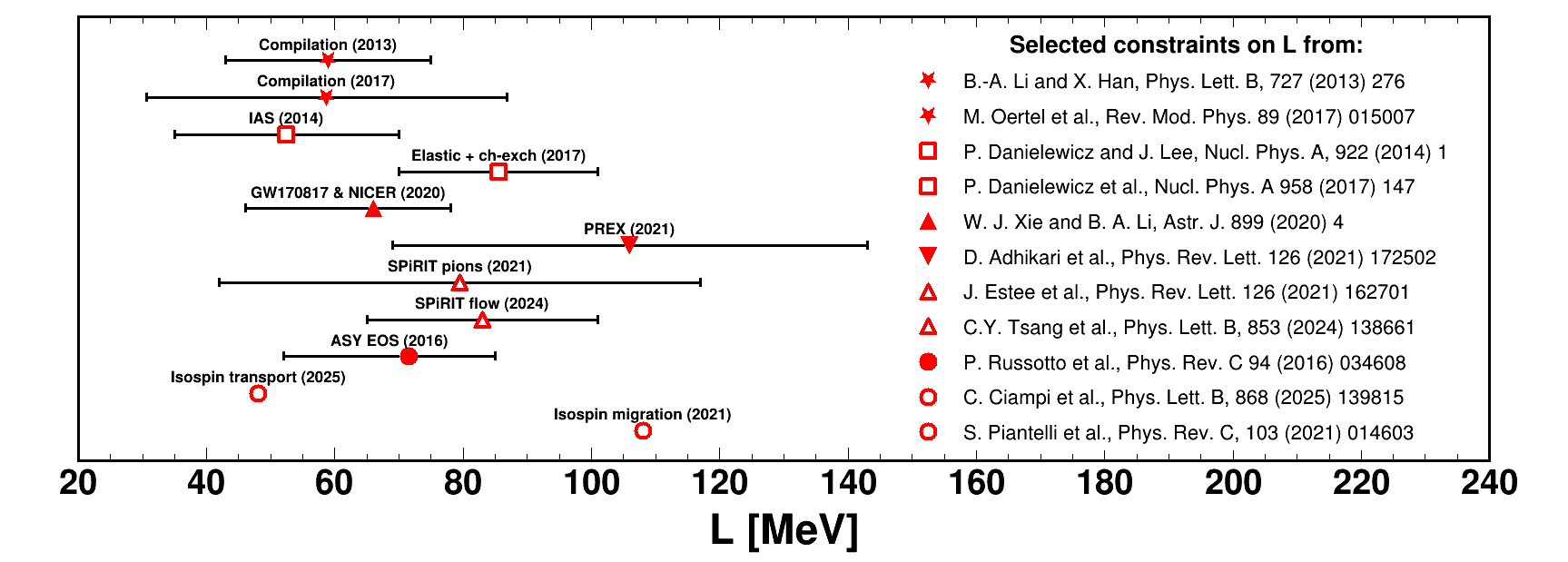}
  \caption{Constraints on the Esym slope parameter from various sources.}
\label{fig_sum} 
\end{figure}

In summary, Fig. \ref{fig_sum} shows a condensed view of the progress in
constraining the L parameter. Recent comparisons of isospin-sensitive
observables with model predictions have yielded new constraints, which are
highlighted in bold in the text and presented as open symbols in
Fig.~\ref{fig_sum}, alongside results from other sources. The most direct link
between isospin and Esym can be established through the IAS and the Lane
potential. Currently, analyses of recent experiments, such as ASY-EOS II and
SPiRIT II, are ongoing, while new data and interpretations are awaited from
neutron and proton removal cross-section measurements. However, care should be
taken when using kaon ratios due to possible enhanced isospin symmetry-breaking
effects, which may require corrections beyond the electromagnetic ones. Notably,
the KRAB detector played an important role in the ASY-EOS II experiment.
Together with future experiments and neutron-star observations, these ongoing
efforts are expected to tighten the constraints further, particularly on the
curvature Ksym and on Esym above twice the saturation density.

\acknowledgments  

The author is grateful to W. Trautmann and B. Zwi\k{e}gli\'{n}ski for their
critical and careful reading of the manuscript and valuable comments and remarks. 

The author used the AI assistant Claude (Opus 5.5, Anthropic) for language editing
and checking of selected passages. All content was reviewed and verified by the
author, who takes full responsibility for the manuscript.

The KRAB system was funded by National Science Centre Poland, NCN, within project
No. 2017/25/B/ST2/02550. We gratefully acknowledge Polish high-performance
computing infrastructure PLGrid (HPC Center: CI TASK) for providing computer 
facilities and support within computational grants PLG/2024/017660 and
PLG/2025/018828. This project has received funding from the European Union's
Horizon 2020 Research and Innovation programme under grant agreement No. 824093 and
from the European Union's Horizon Europe Research and Innovation programme under
Grant Agreement No. 101057511.

\end{document}